\documentclass[prb,aps,twocolumn,showpacs]{revtex4}
\usepackage{epsfig}

\begin{document}

\title{Electronic Structure Studies of Detwinned BaFe$_2$As$_2$ by Photoemission}

\author{Y. K.  Kim$^1$, Hyungju Oh$^1$, Chul Kim$^1$, D. J. Song$^1$, W. S. Jung$^1$, B. Y. Kim$^1$, Hyoung Joon Choi$^1$, C. Kim$^{1,*}$, B. S. Lee$^2$, S. H. Khim$^2$, H. J. Kim$^2$, K. H. Kim$^2$, J. B. Hong$^3$, and Y. S. Kwon$^3$}

\affiliation{$^1$Institute of Physics and Applied Physics, Yonsei University, Seoul 120-749, Korea}

\affiliation{$^2$FPRD, Department of Physics and Astronomy, Seoul National University, Seoul 151-747, Korea}

\affiliation{$^3$Department of Physics, Sungkyunkwan University, Suwon 440-746, Korea}

\date{\today}

\begin{abstract}
We performed angle resolved photoelectron spectroscopy (ARPES) studies on mechanically detwinned BaFe$_2$As$_2$. We observe clear band dispersions and the shapes and characters of the Fermi surfaces are identified. Shapes of the two hole pockets around the $\Gamma$-point are found to be consistent with the Fermi surface topology predicted in the orbital ordered states. Dirac-cone like band dispersions near the $\Gamma$-point are clearly identified as theoretically predicted. At the X-point, split bands remain intact in spite of detwinning, barring twinning origin of the bands. The observed band dispersions are compared with calculated band structures. With a magnetic moment of 0.2 $\mu$$_B$ per iron atom, there is a good agreement between the calculation and experiment. 
\end{abstract}
\maketitle

%Introduction
\section{Introduction}

Recently discovered iron pnictides share important common features with cuprates. Parent compounds have anti-ferromagnetic (AFM) orders\cite{Simon} and AFM orders are suppressed when they are modified by external parameters such as doping or pressure. Superconductivity emerges when the magnetic order is about to be completely suppressed\cite{Simon,Rotter,Sefat}. With these observations, superconductivity in iron pnictides is considered to be related to the magnetic order\cite{Mazin}. Therefore, determining the origin of the magnetic order and its properties can provide important clues to the understanding of the high T$_c$ mechanism in these materials. However, the origin of the magnetic order is not fully understood\cite{Moon} and the size of the magnetic moment remains to be controversial\cite{Mazin2}.

Among various experimental tools, angle resolved photoemission spectroscopy (ARPES) can provide direct information on the electronic structures. For this reason, ARPES experiments have been performed on various iron pnictide compounds%[ARPES papers]
since the discovery of the superconductivity in LaO$_{1-x}$F$_x$FeAs system\cite{Feng,Shen}. From the measured band structures and Fermi surface topology, Fermi surface nesting condition needed for the observed spin density wave (SDW) was examined\cite{Shen}. Based on the comparison between the experimental and calculated bands, the magnetic moment was suggested to be 0.5 $\mu$$_B$ per Fe atom\cite{Shen}. In addition, temperature dependent experiments show band splitting below the magnetic transition temperature as expected from the SDW model\cite{Shen,Feng}.

However, these observations are not without problems. Iron pnictides have structural and magnetic transitions with similar transition temperatures\cite{Simon}. The crystal structure changes from tetragonal to orthorhombic across the transition temperature\cite{Simon}. In the orthorhombic (and magnetic) phase, the system forms twinned crystal and magnetic domains with the axes from two domains orthogonal to each other\cite{Domain}. Existence of such twin domains is not a problem for microscopic probes such as scanning tunneling microscope, but could pose a serious problem for macroscopic tools (such as ARPES and transport measurements) because information from two domains is mixed. If the electronic structure is isotropic, twinning may not have too much effect. Unfortunately, there are several reports that suggest anisotropic electron structures in, for example, BaFe$_2$As$_2$\cite{Dai,TMchuang,Fisher}. So far, most of the measurements have been performed with twinned crystals. So measurements on detwinned (single domain) crystal should be useful.

To make a single domain system, application of an external magnetic field was initially proposed to detwin by aligning the magnetic order\cite{Fishermag}, which unfortunately cannot be applied to ARPES studies. On the other hand, it was recently shown that single domain could be obtained by applying mechanical strain or stress on BaFe$_2$As$_2$\cite{Tanatar,Fisher}. In the single domain transport experiment, it was found that resistivity in BaFe$_2$As$_2$ is quite anisotropic\cite{Tanatar,Fisher}. Because the required external strain to detwin a crystal is relatively low, the method can be used in ARPES experiments. To clarify the issue on the electronic structures in iron-pnictides, we performed ARPES experiments on mechanically detwinned BaFe$_2$As$_2$. Band dispersions with clear sign of detwinning are obtained. The band dispersions are compared with first principles density functional calculation results. Comparison of experimental and calculated dispersions enabled us to extract important information on the electronic structures of BaFe$_2$As$_2$.

\begin{figure}
\centering \epsfxsize=8.5cm \epsfbox{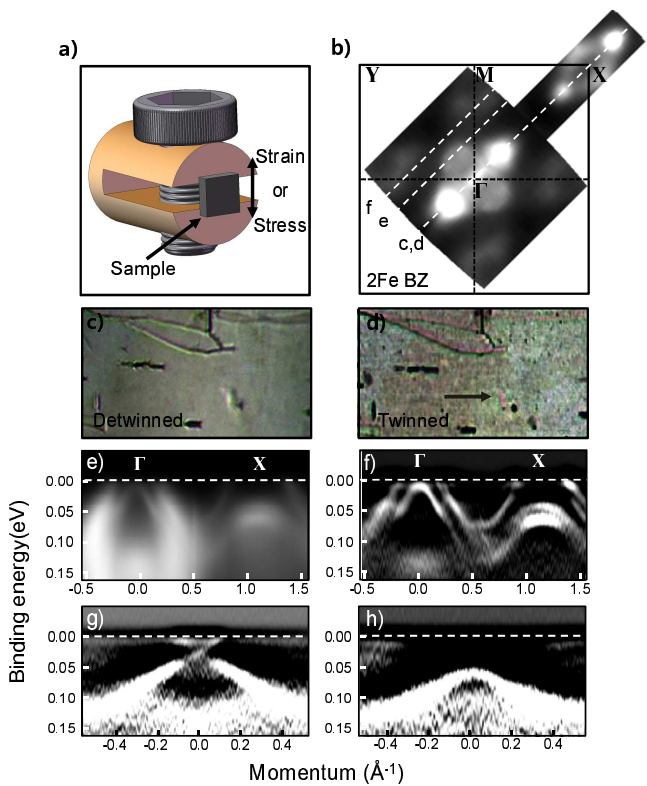} \caption{(Color online) (a) A sample holder designed to apply strain(or stress) to samples. (b) Constant energy map at the Fermi level. (c) Optical microscopy images of detwinned and (d) twinned samples with a polarized light source. images were taken at 88K. The black arrow marks a twin boundary.(e) Raw ARPES data along the $\Gamma$-X direction and (f) its second derivative. (g) and (h) Band dispersions along different cuts as indicated in panel (b).} \label{fig1}
\end{figure}

%Experiment
\section{Experiment}
BaFe$_2$As$_2$ single crystals used in the experiment were grown by self flux method as well as Brigdemann method\cite{Song}. We designed a special sample holder that can apply strain or stress to the sample (see figure 1(a)). To confirm proper detwinning of samples, we took optical microscopy images of samples under the experimental condition with a polarized light source as used in Ref.\cite{Tanatar}(see figure 1(c),(d)). ARPES experiments were performed at the beamline 5-4 of the Stanford Synchrotron Radiation Laboratory equipped with Scienta R4000. The photon energy used in the experiments was 23.7 eV. Energy and momentum resolutions were 16.5 meV and 0.3 degree, respectively. Sample were cleaved at 10K {\it in situ}. Subsequent experiments were also performed at 10 K in a vacuum better than 4$\times$10$^{-11}$ Torr.
Density-functional calculations on the electronic structure of BaFe$_2$As$_2$ are based on {\it ab initio} norm-conserving pseudopotentials\cite{Martins} and the Perdew-Burke-Ernzerhof-type generalized gradient approximation\cite{Perdew}, as implemented in the SIESTA package\cite{Soler}. Experimental lattice constants and atomic positions in the low temperature antiferromagnetic phase\cite{Huang} are used in the calculations except for the As height which is shifted by 0.07 $\AA$ further away from the Fe layer. Constraint is imposed on the electron density to make the magnetic moment be 0.2 $\mu$$_B$ at each Fe atom.

%Results
\section{Results}
%Fig. 1 : general information(fermi surface topology)
In figure 1(b), we plot the Fermi surface map from a detwinned sample. With an inner potential of $V_0=14$ eV from the literature\cite{Brouet}, the data is for $k_z=0$. General features of the Fermi surface do not differ drastically from those of twinned samples. Near the M-point, even though weak, Fermi surface pockets are observed. We attribute these features to surface states due to surface reconstruction observed in scanning tunneling microscopy studies\cite{Nas,Mas}.

%Fig. 1-2 : general information(band dispersion)
To see more detailed electronic structure information, we plot ARPES data along several different momentum space cuts in figure 1(e)-(h). The directions of the cuts are shown in figure 1(b). Figure 1(e) shows raw data along the $\Gamma$-X high symmetry direction and figure 1 (f) is its second derivative. From these plots, we see clear dispersive features and can determine band dispersions. The band dispersions along the $\Gamma$-X in figure 1(f) are not significantly different from those obtained from twin domain samples\cite{Shen}.
Most notably, split bands at the X-point that appear below the magnetic transition still exist after detwinning. They were initially interpreted as being due to exchange splitting\cite{Feng} but later were argued to be from different domains. Our observation of the split bands after detwinning reveals that they are genuine features of magnetically ordered state. On the other hand, away from the $\Gamma$-point, the band dispersion appears quite different (figure 1(g) and (h)). First of all, we see only two crossing bands in figure 1(g), of which the dispersion looks similar to that of a Dirac cone observed in graphene and topological insulators. The fact that other bands observed in twin domain samples\cite{Shen} are not seen confirms proper detwinning of our samples.
 On the other hand, figure 1 (h) shows the band dispersion parallel to the $\Gamma$-X direction and weak surface states pockets mentioned above.

%%Fig. 2 : about Fermi pocket
So far, we have focused on the band dispersion. Now let us move on and determine the characters of Fermi surface pockets as they may provide us with information on the transport properties of the system. Normally, two different methods can be used to investigate the characters of Fermi surface pockets. One way is to consider the change in the pocket size in the constant energy map as a function of the binding energy. The other way is to determine it with the band dispersion of a certain pocket. As several pockets are closely located near the $\Gamma$-point and the overall features are too broad to distinguish each pocket, we use the latter method to determine the character of each Fermi surface pocket.

\begin{figure}
\centering \epsfxsize=8.5cm \epsfbox{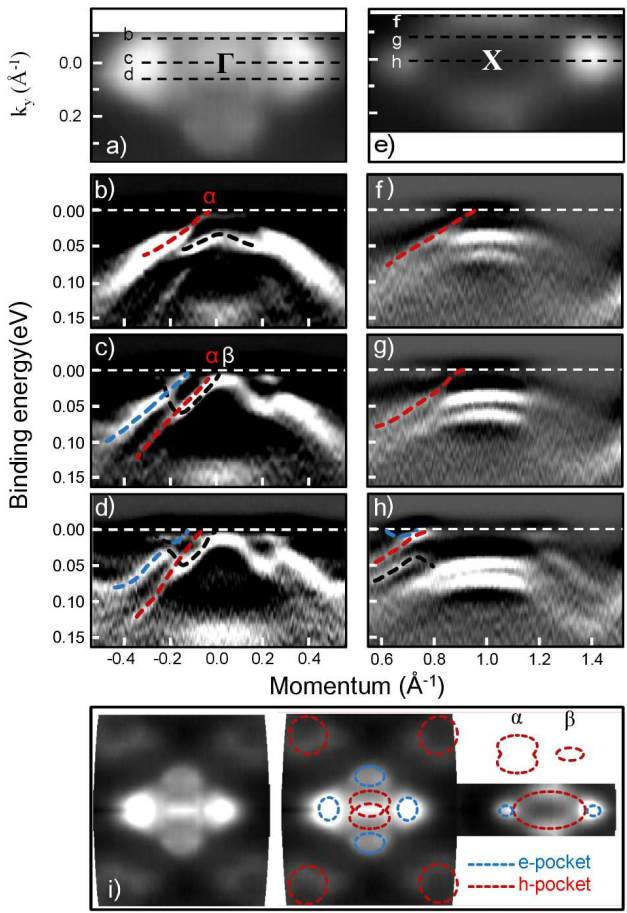} \caption{(Color online) (a) Fermi surface map around the $\Gamma$-point. (b) to (d) Second derivatives of ARPES data along the cuts indicated by the dashed lines in panel (a). (e) Fermi surface map around the X-point. (f) to (h) Band dispersions from second derivatives of the data along the cuts indicated by the dashed lines in panel (e). (i) Characters of the Fermi surface pockets. Red and blue circles indicate hole and electron pockets, respectively. The shape of the two hole pockets around the $\Gamma$-point are shown in the inset. Definition of the angle ($\theta$) used in figure 4 is also shown.}
\label{fig2}
\end{figure}

First, we focus on the pockets around the $\Gamma$-point. The weak pockets which from a surface state and the Dirac cone pockets are already mentioned and were determined to be hole and electron pockets, respectively. On the other hand, other pockets around the $\Gamma$-point are closely located and thus are more difficult to determine the character. To resolve the problem, we plot detailed cuts in figure 2. In figures 2(b)-(d), second derivatives of ARPES data are plotted to see the band dispersions around the $\Gamma$-point. The cuts in the momentum space are indicated in the figure 2(a). In figure 2(b), we can see a band that crosses the Fermi level at k$_F$ = -0.098 $\AA^{-1}$ (red dashed line, labeled as $\alpha$ band) which forms a hole like $\alpha$ pocket around the $\Gamma$-point. In addition, we also see another band (black dashed line, $\beta$ band) at higher binding energy. To find the detailed shape of the pockets, we trace the Fermi momentum $k_F$ change around the $\Gamma$-point.

Along the $\Gamma$-X high symmetry line, k$_F$ value is relatively small (-0.053 $\AA^{-1}$, figure 2(c)) and becomes larger (-0.070 $\AA^{-1}$) when the cut moves away from the $\Gamma$-X line (figure 2(d)). The shape of the $\alpha$ pocket found in this way is shown in the inset in figure 2(i). It is a hole pocket with a form of deformed circle.

On the other hand, the $\beta$ band crosses the Fermi level only near the $\Gamma$-point and forms an ellipsoidal hole pocket($\beta$ pocket) as also illustrated in the inset. The properties of these two pockets are consistent with the prediction when there is orbital ordering in the system\cite{Tom}. Around these two pockets, there are four electron pockets. The shapes and characters of Fermi pockets around the $\Gamma$-points are summarized in figure 2(i).

We now look at the pockets near the X-point. Figure 2(e) plots a Fermi surface map near the X-point while figures 2(f)-(h) show the ARPES data along the cuts indicated in figure 2(e). Away from the X-point, there are two electron-like pockets on the $\Gamma$-X high symmetry line (figure 2(h)). Between these two pockets, there is a large hole-like pocket for which the band disperses away from the X-point as the binding energy increases, making it a hole pocket.

\begin{figure}
\centering \epsfxsize=8.5 cm \epsfbox{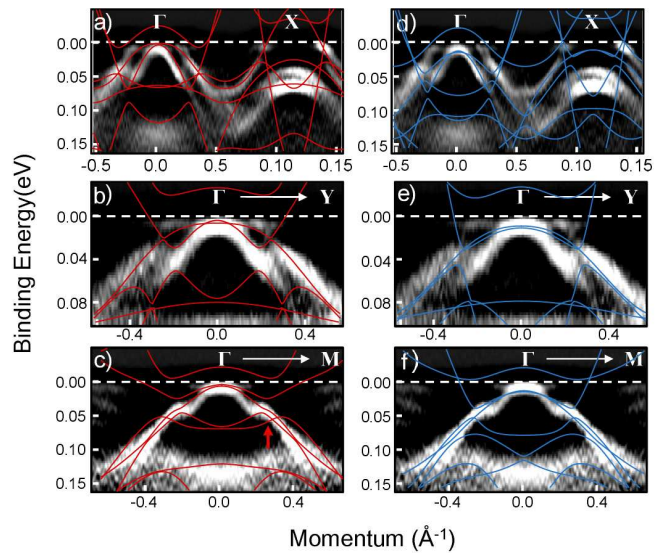} \caption{(Color online)(a)-(c) Band dispersions along three different high-symmetry lines, $\Gamma$-X, $\Gamma$-Y, and $\Gamma$-M overlaid with calculated bands with magnetic moment of 0.2 $\mu$$_B$. The As height was adjusted for the best fit and is lager by 0.070 $\AA^{-1}$ than the experimentally obtained value. (d)-(f) Bands are calculated with the experimentally obtained As height. Calculated bands are renormalized by a factor of 3 and Fermi level was shifted by 25 meV.}
\label{fig3}
\end{figure}

%Fig. 3 : +LDA - magnetic moment
Once the experimental dispersions are determined, it is important to compare them with the calculated band structure. By comparing them, one may extract useful physical quantities, especially the effective magnetic moment of the ordered state. The size of the magnetic moment is under debate due to the mismatch between the predicted and observed values\cite{Toyweak}. Latest value obtained by comparing the experimental and calculated band structures is 0.5 $\mu$$_B$ but it was based on the experimental band structure from twinned samples\cite{Shen}. In figure 3, we plot ARPES data as well as calculated band dispersions. We calculated the band structure with various values of magnetic moments, including 0.5 $\mu$$_B$. The best match between experimental and calculated dispersions was given when we set the magnetic moment of the magnetically ordered state to be 0.2 $\mu$$_B$ (figure 3(a) to (c)). As a side note, we not only adjusted the magnetic moment but also the arsenic height ($\Delta z_{As}=0.07\AA$ higher than experimentally measured height\cite{Huang}) for a better match.

The magnetic moment we obtained is quite small compared to the previously used value of 0.5 $\mu$$_B$\cite{Shen} but is close to a recently considered value in theoretical studies\cite{Toyweak,Valen}. It is also consistent with a recently suggested value of 0.19 $\mu$$_B$ from single domain ARPES  data from CaFe$_2$As$_2$\cite{Dan}. This probably means that the correlation between electron and magnetic order in the system is not strong, that is, the band calculation overestimates the electron-magnetic order interaction.

%Fig. 3 : +LDA - Important features
Figure 3(a) to (c) compare experimental and calculated band dispersions along three different high-symmetry lines, $\Gamma$-X, $\Gamma$-Y and $\Gamma$-M. Even though the match between them is quite good, there are few features to be noted. In panel (a), two parallel parabolic bands at the $\Gamma$-point close to the Fermi energy match the experimental dispersions. More remarkably, two split bands at the X-point at the binding energies of 42 and 64 meV are reasonably reproduced.

As mentioned earlier, the origin of these bands were under debate\cite{Feng,Shen}. This issue is now clarified with the detwinned data. Band calculation shows an excellent match with the experimental data, confirming them to be genuine features of magnetically ordered state. Lastly, a double bent feature in the band dispersion near the $\Gamma$-point (marked by an arrow in (c)) is not a single band but can be explained by existence of multiple bands. Some parts of the bands are not seen due to the orbital characters and accompanying selection rules.

%Fig. 3 : +LDA - Asenic height
In figures 3(d) to (f), we plot calculated bands with the experimentally measured arsenic height\cite{Huang}. The As height effect on the band structure is rather drastic. The features mentioned in the previous paragraph such as the double bent feature cannot be explained by the calculated band structure. The effect is especially strong near the X-point where the two parallel parabolic bands are now separated by about 0.1 eV, much larger than experimental value.

\begin{figure}
\centering \epsfxsize=8.5cm \epsfbox{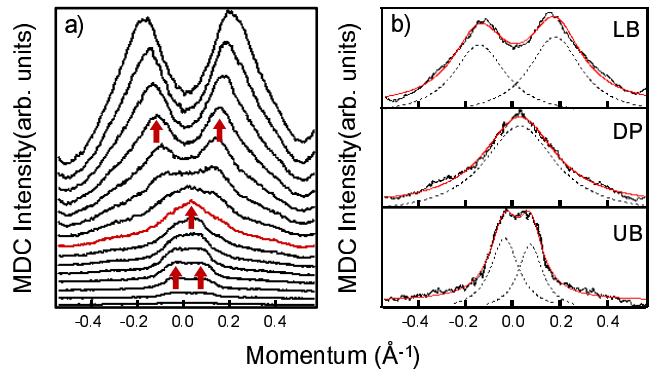} \caption{(Color online) (a) MDCs with different binding energies. (b) Fitting results for MDCs from above (UB), at (DP) and below (LB) the Dirac point.}
\label{fig4}
\end{figure}

%Fig. 4 : Dirac cone
Lastly, we discuss the Dirac cone band dispersion shown in figure 1(e). There are theoretical results predicting that Dirac cone-like band dispersions should appear in the electronic structure of iron pnictide compounds\cite{DHLee,Toyama}. Recently, there has been a report on observation of a Dirac cone dispersion in twinned samples by ARPES experiment\cite{Takahashi}. At this point, we need to point out that the Dirac cone presented in figure 1(e) is along the $\Gamma$-Y direction and thus different from the Dirac cone along the $\Gamma$-X direction. However, both Dirac cones have the same origin in that the original and folded bands cross each other without hybridization gap because they have different parities\cite{DHLee}.

To clearly show that the band dispersion is really Dirac cone-like, we plot momentum distribute curves (MDCs) at various binding energies in figure 4(a). The peak positions indicated by the arrows already show that the two bands cross each other. To further confirm it, we took three of them and plot them in figure 4(b) with fitting results. We observe that the two bands cross each other at the Dirac point without an hybridization gap. Such crossing band feature exists in the calculated band structure shown in figure 3(b). From these observations, we conclude that Dirac cone-like bands exist (both along the $\Gamma$-X and $\Gamma$-Y directions) in the electronic structure of BaFe$_2$As$_2$. For the $\Gamma$-Y direction, the Dirac point is located at a binding energy of 23 meV which is consistent with the theoretically predicted value\cite{Toyama}.

Existence of such a Dirac band in iron pnictide compounds provides us an important clue to understanding the origin of magnetism in iron pnictides. The origin of the magnetism in iron pnictides is still under debate because no sign of an SDW gap has been observed in ARPES experiments. However, it was claimed that there should be no hybridization gap if the parity of the folded band has a parity opposite to that of the original band\cite{DHLee}. With the same parity argument, it was predicted that there should be Dirac cone dispersions in iron pnictide compounds. Therefore, existence of Dirac cones in the experimental ARPES data reveals that the parity of a band is a good quantum number in the system and plays an important role in determining the electronic structure.

\section{Conclusion}
%Conclusion
In conclusion, we performed ARPES experiments on mechanically detwinned BaFe$_2$As$_2$ and obtained the experimental band structures. At the M-point, a surface state hole pocket is observed and around the $\Gamma$-point Dirac band dispersions are observed. We identify the Fermi surface topology around the $\Gamma$- as well as X-points. We also find that the split bands at the X-point are a genuine features of magnetic phase, not an artifact due to twinning. A magnetic moment of 0.2 $\mu$$_B$ gives the band structure that best matches the experimental dispersions. Electronic structure is found to be very sensitive to the arsenic height as known before. Existence of Dirac cones reveals that the parity of a band could play an important role and thus should be properly considered in the theories for iron pnictide systems.

%Note added in proof
\emph{Note added in proof$:$} M. Yi \emph{et al.}\cite{Yi} have also performed similar experiment and it was posted recently after the our submission.

%Acknowledgments
\begin{acknowledgments}
Authors would like to thank J. H. Han, S. R. Park and T. Tohyama for
helpful discussions. This work was supported by the KICOS in No. K20602000008 and by Mid-career Researcher Program through NRF grant funded by the MEST (No. 2010-0018092). Computational part was supported by the NRF of Korea (Grant 2009-0081204) and computational resources have been provided by KISTI Supercomputing Center (Project No. KSC-2008-S02-0004). SSRL is operated by the DOEs Office of BES.
\end{acknowledgments}

\end{document}